\documentclass[preprint,pre,aps,floatfix,nofootinbib]{revtex4}

\usepackage{graphicx}

\newcommand{\be}{\begin{equation}}
\newcommand{\ee}{\end{equation}}

\newcommand{\fig}[1]{Fig.~\ref{#1}}

\newcommand{\eq}[1]{Eq.~(\ref{#1})}
\newcommand{\eqs}[1]{Eqs.~(\ref{#1})}

\newcommand{\bd}{{\rm b}}
\newcommand{\piad}{\pi_{\rm ad}}
\newcommand{\pieps}{(\pi_{\rm ad}/\epsilon)}

\newcommand{\dd}{{\rm d}}

\def\siml{\mathrel{\mathchoice {\vcenter{\offinterlineskip\halign{\hfil
$\displaystyle##$\hfil\cr<\cr\sim\cr}}}
{\vcenter{\offinterlineskip\halign{\hfil$\textstyle##$\hfil\cr
<\cr\sim\cr}}}
{\vcenter{\offinterlineskip\halign{\hfil$\scriptstyle##$\hfil\cr
<\cr\sim\cr}}}
{\vcenter{\offinterlineskip\halign{\hfil$\scriptscriptstyle##$\hfil\cr
<\cr\sim\cr}}}}}

\def\simg{\mathrel{\mathchoice {\vcenter{\offinterlineskip\halign{\hfil
$\displaystyle##$\hfil\cr>\cr\sim\cr}}}
{\vcenter{\offinterlineskip\halign{\hfil$\textstyle##$\hfil\cr
>\cr\sim\cr}}}
{\vcenter{\offinterlineskip\halign{\hfil$\scriptstyle##$\hfil\cr
<\cr\sim\cr}}}
{\vcenter{\offinterlineskip\halign{\hfil$\scriptscriptstyle##$\hfil\cr
<\cr\sim\cr}}}}}

\begin{document}

\title[Cooperative cargo transport]{Cooperative Cargo Transport by \\
Several Molecular Motors}

\author{Stefan Klumpp and Reinhard Lipowsky}

\address{Max Planck Institute of Colloids and Interfaces, Science Park Golm, 
14424 Potsdam, Germany}

\begin{abstract}
The transport of cargo particles which are pulled by several
molecular motors in a cooperative manner is studied theoretically.
The transport properties depend primarily on the maximal
number, $N$, of motor molecules that may pull simultaneously
on the cargo particle.
Since each motor  must unbind  from the filament after a finite 
number of steps but can also rebind to it again, the  actual  number of pulling
motors is not constant but varies with time between zero 
and $N$. An increase in the maximal  number
$N$ leads to a strong increase of the
average walking distance (or run length) of the cargo particle.
If the cargo is pulled by up to $N$ kinesin motors, e.g., the walking
distance is estimated to be $5^{N-1}/N$ micrometers which implies
that seven or eight kinesin molecules are sufficient to attain an
average  walking distance in the centimeter range.
If the cargo particle is pulled  against an external load force,  
this force is shared between the motors which provides
a nontrivial  motor-motor coupling and 
a generic mechanism for nonlinear force-velocity relationships. 
With increasing load force, the probability distribution of the 
instantenous velocity is shifted towards smaller values, becomes
broader, and develops several peaks. Our theory is 
consistent with available experimental data and makes
quantitative predictions that are accessible to systematic in vitro
experiments.

\end{abstract}

\maketitle

\section{Introduction}

Cytoskeletal motors which perform active movements along cytoskeletal
filaments drive the long-range transport of vesicles, organelles, and
other types of cargo in biological cells.  In the following, we will
consider \emph{processive} motors which can complete many
chemo-mechanical cycles while remaining bound to the filaments.
During the last decade, the properties of \emph{single} processive
motors such as kinesin on microtubules and myosin V on actin filaments
have been characterized in some detail using \emph{in vitro} motility
assays and novel experimental techniques for the visualization and
manipulation of single molecules \cite{Howard2001,Schliwa2003}.
However, \emph{in vivo}, force generation and transport is typically
performed by \emph{several} motor molecules in a cooperative fashion
as revealed by electron microscopy
\cite{Miller_Lasek1985,Ashkin__Schliwa1990} and by tracking of the
cargo particles with optical methods
\cite{Gross__Wieschaus2002,Hill__Holzwarth2004,Kural__Selvin2005}.  It
has also been found that some cargo particles bind different types of
motors simultaneously so that these particles can reverse their
direction of motion along microtubules
\cite{Gross__Wieschaus2002,Kural__Selvin2005} or switch from
microtubules to actin filaments \cite{Kuznetsov__Weiss1992}.

The force generated by a single cytoskeletal motor is rather small and
of the order of a few piconewtons. Larger forces can be generated if
several motors pull on the same cargo.  This is necessary, e.g., for
the fast transport of large organelles through the cytoplasm which is
a highly viscous medium \cite{Luby-Phelps2000}. Likewise, large forces
arising from many motors are also required for specific motor
functions such as the extraction of membrane tubes from vesicles
\cite{Koster__Dogterom2003,Leduc__Prost2004}.

Another important consequence of the cooperative action of several
motors is that it increases the walking distance (or run length) of
the cargo particles. Since the binding energy of such a cargo particle
is necessarily finite, it can be overcome by thermal fluctuations
which are ubiquitous in cells.  If the cargo particle is pulled by a
single processive motor, its walking distance is typically of the
order of one micrometer \cite{Lipowsky__Nieuwenhuizen2001}.  If the
cargo particle is pulled by several motors, the walking distance is
strongly increased since the cargo continues to move along the
filament unless all motors unbind simultaneously.  In addition, as
long as the cargo particle is still connected to the filament by at
least one motor, all unbound motors can rebind rather fast, because
they are prevented from diffusing away from the filament.  It has also
been shown using in vitro motility assays that cargo particles pulled
by many motors can switch tracks and move along several filaments at
the same time, so that huge walking distances can be achieved which
exceed the length of a single filament \cite{Boehm__Unger2001}.

In this article, we study these cooperative transport phenomena from
the theoretical point of view. First, we introduce a generic
transition rate model for the transport of cargo particles, which are
pulled by up to $N$ motors, and obtain general expressions for the
 average number of pulling motors,  for the 
average velocity of the bound cargo particle, for its effective
unbinding rate, and for the distribution of its walking distances.
Next, we focus on the case of cargo particles with a dilute motor
coverage, which should be directly applicable to typical bead assays.
In the absence of an external load force, we obtain an explicit
expression for the average walking distance of the cargo particles.
Using this expression for particles that are pulled by up to $N$
kinesin motors, we estimate the walking distance to grow as
$5^{N-1}/N$.  We also calculate the distribution of the walking
distances which is found to exhibit a tail with an extended plateau
region for $N \ge 3$.

An external load force leads to a nontrivial coupling between the
different motors because the unbinding rates of the motors increase
with increasing force.  As a consequence, the average number of bound
motors decreases as the load force is increased which provides a
generic mechanism for {\em nonlinear} force-velocity relationships. We
argue that the motor transport becomes ineffective at a critical force
for which the average walking distance becomes comparable with the
step size of a single motor.  For $N \ge 2$, this critical force is
found to be small compared to the maximal stall force which can be
sustained by $N$ motors.  Finally, we calculate the probability
distribution of the instantaneous velocity of the bound cargo
particle. As the load force is increased, this velocity distribution
is shifted towards smaller values, becomes broader, and develops
several peaks.

We will focus on the transport by kinesin motors, which pull cargo
particles along microtubules, since, in this case, all input
parameters for our theory have been determined experimentally, but our
analysis is rather general and can be applied to other types of
cytoskeletal motors as well.  All experimental data that are available
for cargo transport by several kinesin motors are consistent with our
theoretical results.

\section{Model and General Solution}
\label{sec:theory}

\begin{figure}[h]
   \begin{center}
     \vspace*{0.5cm}
     \leavevmode
     \includegraphics[angle=0,width=.9\columnwidth]{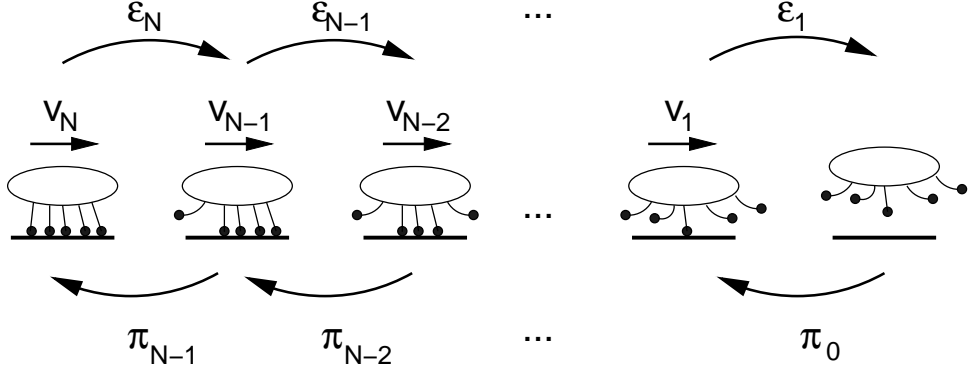}
     \caption{A cargo particle is transported cooperatively by  $N$ 
       molecular motors along a filament. The motors are firmly attached
       to the cargo but unbind from and rebind to the filament. 
       Each state  of the system, denoted by $|n\rangle$,  is
       characterized by the number $n$ of bound motors that pull 
       on the cargo particle. 
        The latter number can vary between $n=N$
        (on the left) and $n=0$ (on the right). 
        In state $|n\rangle$, the cargo particle has 
        velocity $v_n$, a motor
       unbinds from the filament with rate $\epsilon_n$, and an
       additional motor binds to the filament with rate $\pi_n$.}
     \label{fig:NMot1Cargo}
   \end{center}
\end{figure}

{\bf Transition rate model. } 
We consider cargo particles which are pulled by $N$
motors, see \fig{fig:NMot1Cargo}. These motors are irreversibly
attached to the cargo particle, but can bind to and unbind from the
filament along which they move. Thus, the number $n$ of motor
molecules that are bound to the filament can vary between $n=0$ and
$n=N$. We will distinguish $N+1$ different states of the cargo
particle corresponding to the unbound state with $n=0$ and to $N$
bound states with $n=1,2,\dots N$. Each of these bound states contains
$N!/(N-n)!n!$ substates corresponding to the different combinations of
connecting $n$ motor molecules to the filament.  If the cargo particle
is linked to the filament through $n$ motors, it moves with velocity
$v_n$. Unbinding of a motor from the filament and binding of an
additional motor to the filament occur with rates $\epsilon_n$ and
$\pi_n$, respectively.

We first derive general expressions for the transport properties of
the cargo particles pulled  by $N$ motors without specifying how the
rates $\epsilon_n$ and $\pi_n$ and the velocities $v_n$ depend on the
number $n$ of bound motors. We derive the distributions of the number
of bound motors, of the binding times and of the walking distances
from which we then obtain the effective unbinding rate, the average
walking distance, and the average velocity. All these quantities can
be directly measured by particle tracking both in vivo and in vitro.

{\bf Distribution of the number of bound motors. } We first calculate
the distribution of the number of bound motors. We denote by $P_n$ the
probability that the cargo particle is in state $|n\rangle$, i.e.\ bound to
the filament by $n$ motors.  These probabilities satisfy the master
equation
\begin{equation}
  \frac{\partial}{\partial t} P_n=\epsilon_{n+1}P_{n+1}+\pi_{n-1}P_{n-1}-
  (\epsilon_n+\pi_n)P_n.
\end{equation}
We are interested in the transport properties of bound cargo
particles. Since all movements of bound cargo particles begin and end
with $n=0$, every step from state $|n\rangle$ to $|n+1\rangle$ implies a backward step
at some later time. To determine the transport properties of the bound
cargo particles, we can therefore focus on the stationary solution of
the master equation which is characterized by
\begin{equation}\label{eq:stat}
  \epsilon_{n+1}P_{n+1}=\pi_n P_n
\end{equation}
for $0\leq n\leq N-1$.

Expressing subsequently all $P_n$ in terms of $P_0$ and using the
normalization $\sum_{n=0}^{N}P_n=1$, we obtain
\begin{equation}
  \label{eq:Pk}
  P_0=\left[1+\sum_{n=0}^{N-1}\prod_{i=0}^{n}\frac{\pi_i}{\epsilon_{i+1}}\right]^{-1} \qquad{\rm and}\qquad P_{n}= P_0
\prod_{i=0}^{n-1}\frac{\pi_i}{\epsilon_{i+1}}. 
\end{equation}
To determine the transport properties of cargo particles bound to the
filament, we normalize these probabilities with respect to the bound
states, i.e.\ we consider the
the probabilities $P_n/(1-P_0)$  that a bound cargo particle
is bound to the filament by $n$ motors.
For example, the average number of bound motors
is given by
\be
N_\bd=\sum_{n=1}^{N}  \, n \, P_n/(1-P_0) . 
\label{NBoundMotors}
\ee

\vspace*{0.4cm}
{\bf Average velocity.} The distribution of the number of bound
motors as given by \eq{eq:Pk} implies the distribution of velocities
of the cargo particle moving along the filament
\begin{equation}
  \label{eq:histv}
  P(v)=\sum_{n=1}^{N}\delta(v-v_n)\frac{P_n}{1-P_0}.
\end{equation}
The latter quantity can be determined experimentally as the histogram
of velocities averaged over short time intervals.
The average velocity of the cargo particle moving along the filament
is then given by
\begin{equation}
  \label{eq:v_eff}
  v_{\rm eff}=\sum_{n=1}^{N} v_n \frac{P_n}{1-P_0}.   
\end{equation}
If the velocity of the cargo particle is independent of the number of
bound motors, $v_n=v$, the effective velocity is equal to the
single-motor velocity $v$.

{\bf Effective unbinding rate. } Finally, the distribution of the
number of bound motors implies also an explicit expression for the
effective detachment or unbinding rate. In the stationary state,
the effective binding and unbinding rates, $\pi_{\rm eff}$ and
$\epsilon_{\rm eff}$, fulfill the simple relation
\begin{equation}
  \epsilon_{\rm eff}(1-P_0)=\pi_{\rm eff}P_0
\end{equation}
where $(1-P_0)$ 
is again the probability that the cargo particle is bound to the
filament through at least one motor. The effective binding rate is
given by $\pi_{\rm eff}=\pi_0$ since the cargo--filament link is
established as soon as one motor binds to the filament, so that
$\epsilon_{\rm eff}=\pi_0 P_0/(1-P_0)$.\footnote{This definition is
  equivalent to defining the effective unbinding rate as
  $\epsilon_{\rm eff}=\epsilon_1 P_1/(1-P_0)$, i.e.\ as the unbinding
  rate of the last bound motor times the probability that a cargo
  particle bound to the filament is linked to this filament by a
  single motor.}

With the distribution of the number of bound motors as given by
\eq{eq:Pk}, we obtain
\begin{equation}
  \label{epsilon_eff}
  \epsilon_{\rm eff}= \epsilon_1  \left( 1+\sum_{n=1}^{N-1}\prod_{i=1}^n
  \frac{\pi_i}{\epsilon_{i+1}} \right)^{-1} , 
\end{equation}
where the summation now starts with $n=1$.
For $N=2$ motors, this result 
reduces to $\epsilon_{\rm eff}=\epsilon_1/(1+\pi_1/\epsilon_{2})$.
An alternative derivation of (\ref{epsilon_eff}) based on first
passage times is
presented in part A.1 of the Supporting Information.

{\bf Distributions of binding times and walking distances. } The
effective unbinding rate as given by \eq{epsilon_eff} determines only
the average time that the cargo particle is bound to the filament. The
actual binding time $\Delta t_\bd$ of the cargo particles is, however,
a stochastic quantity which is governed by a certain probability
distribution $\tilde\psi_N(\Delta t_\bd)$.

This probability distribution governs the passage from the state with
one motor connecting the cargo to the filament at time $t$
(immediately after binding) to the unbound state at time $t+\Delta
t_\bd$.
This distribution can be obtained by solving a recursion relation 
as shown in part A.2 of the Supporting Information. 
The general solution is a sum of exponentials,
\begin{equation}\label{eq:psi_t_genSol}
  \tilde\psi_{N}(\Delta t_\bd)=\sum_{i=1}^{N}e^{-z_i \Delta t_\bd}\ {\rm Res}(-z_i),
\end{equation}
where the scales $-z_i$ of the exponentials and the prefactors ${\rm
  Res}(-z_i)$ are the poles and the corresponding residues,
respectively, of a fraction
of polynomials which is given in the Supporting Information. 
The time scales and prefactors are functions of the binding and
unbinding rates and should not be considered as independent fit
parameters when analyzing experimental data.

The distribution of the walking distances, 
$\psi_{N} (\Delta x_\bd)$, is obtained from the distribution of binding
times by substituting $\Delta t_\bd$ by $\Delta x_\bd$, $\epsilon_n$
by $\epsilon_n/v_n$, and $\pi_n$ by $\pi_n/v_n$, i.e.,  by expressing
the rates in units of  (inverse) distance traveled rather than in
units of inverse time. 
The distribution $\psi_{N} (\Delta x_\bd)$
 is therefore also given by a sum of $N$
exponentials as in  
\eq{eq:psi_t_genSol}
 and has the general form
\begin{equation}
\label{eq:psi_x_genSol}
\psi_{N} (\Delta x_\bd)=\sum_{i=1}^{N}e^{-z'_i \Delta x_\bd}\ {\rm Res}(-z'_i).
\end{equation}
The same substitution leads to an explicit expression for the average walking
distance $\langle\Delta x_\bd\rangle$ as given by
\begin{equation}
  \label{eq:deltax_N}
  \langle\Delta x_\bd\rangle=\frac{v_1}{\epsilon_1}\left[1+\sum_{n=1}^{N-1}
\prod_{i=1}^n \frac{v_{i+1}\pi_i}{v_i\epsilon_{i+1}}\right]
\end{equation}
which again applies to a cargo particle pulled by $N$ motors.

\section{Results}

{\bf Cargo particles with dilute motor coverage. }
Let us now consider specific examples and specify the dependence of the
rates $\pi_n$ and $\epsilon_n$ and of the velocity $v_n$ on the
number $n$ of bound motors.  First, we consider the case where
the cargo particle is transported by  $N$ motor molecules which have 
well-separated anchor points on the particle surface and which, 
thus, do not experience mutual interactions.  
In the absence of an external load force, the parameters 
$\epsilon_n$, $\pi_n$ and $v_n$  are then given by
\begin{equation}
\label{eq:rate_NoWW}
  \epsilon_n=n \, \epsilon,\qquad\pi_n=(N-n) \, \piad,\qquad{\rm and}\qquad v_n=v, 
\end{equation}
where $\epsilon$, $\piad$, and $v$ are the unbinding rate, the binding
rate, and the velocity of a {\it single} motor,
respectively.\footnote{In the present context, the binding rate $\piad$ 
corresponds to a motor which
  remains close to the filament because of the presence of the other
  motors connecting the cargo to the filament. In general, one should
  use $\pi_n=(N-n)\piad$ only for $n\geq 1$ and specify $\pi_0$
  separately in order to account for the diffusion of the completely unbound
  cargo.  The transport properties of the \emph{bound} cargo particle
  are however independent of the choice for $\pi_0$.}

In the following, we use parameter values for kinesin motors as summarized
in Table \ref{tab:param} to determine numerical results, but the
general expressions can also be applied to other types of motors. Our
model with rates as specified by \eq{eq:rate_NoWW} has three
parameters which can be determined from the studies of single motor
molecules: the velocity $v$, the unbinding rate $\epsilon$, and the
binding rate $\piad$.  The first two quantities have been measured for
many types of motors. For kinesin, the velocity is about $1\mu{\rm
  m}/$s and the unbinding rate is about $1/$s
\cite{Block__Schnapp1990,Vale__Yanagida1996}. The binding rate is more
difficult to measure.  If $\piad$ is regarded as an unknown quantity,
our results for the effective unbinding rate or the distribution of
walking distances can be used to determine $\piad$ experimentally.
Here, we use $\piad\simeq 5/$s as measured for kinesins linking a
membrane tube (which acts as the cargo particle) to a microtubule
\cite{Leduc__Prost2004}.

For the parameters as specified by (\ref{eq:rate_NoWW}), the general 
expression  (\ref{NBoundMotors}) for the
 average number of bound motors implies the explicit relation 
  \be
 N_\bd = \frac{ \pieps \left[1 + \pieps \right]^{N-1}}{  
  [1+\pieps]^N - 1 } \,  N   
 \ee
which implies the simple asymptotic behavior
 $N_\bd \approx   \frac{  \pieps}{ 1 + \pieps} \, N $ for large $N$. 

Likewise, the general expression 
 (\ref{eq:deltax_N}) for the average walking distance 
 $\langle\Delta x_\bd\rangle$ can be
evaluated analytically which leads to
\begin{equation}
  \label{eq:deltax_N_noWW}
  \langle\Delta x_\bd\rangle=\frac{v}{\epsilon_{\rm eff}}
  =\frac{v}{N\piad}\left[\left(1+\frac{\piad}{\epsilon}\right)^N-1\right].
\end{equation}
For strongly binding motors with $\piad/\epsilon\gg 1$, the walking
distance behaves as $\langle\Delta x_\bd\rangle\approx
\frac{v}{N\epsilon}(\piad/\epsilon)^{N-1}$ and essentially increases
exponentially with increasing number of motors. For weakly binding
motors, $\langle\Delta x_\bd\rangle\approx
(v/\epsilon)[1+\frac{N-1}{2}\frac{\piad}{\epsilon}]$, where the
leading term $v/\epsilon$ corresponds to the walking distance of a
single motor.

Kinesin binds rather strongly with $\piad/\epsilon\simeq 5$, so that
the average walking distance, which is 1 $\mu$m for a single motor,
increases quickly with $N$ and is 3.5~$\mu$m, 14~$\mu$m, 65~$\mu$m,
and 311~$\mu$m for cargoes pulled by 2, 3, 4, or 5 motors,
respectively.
These large walking distances exceed the length of a single
microtubule, but can still be realized if several microtubules are
aligned in a parallel and isopolar fashion, so that, via unbinding and
rebinding, the motors can step from one microtubule to another. Such
an organization of microtubules is typical for axons
\cite{Goldstein_Yang2000} and has also been engineered in vitro
\cite{Boehm__Unger2001}.

\begin{figure}[h]
   \begin{center}
     \vspace*{0.2cm}
     \leavevmode
     \includegraphics[angle=0,width=0.9\columnwidth]{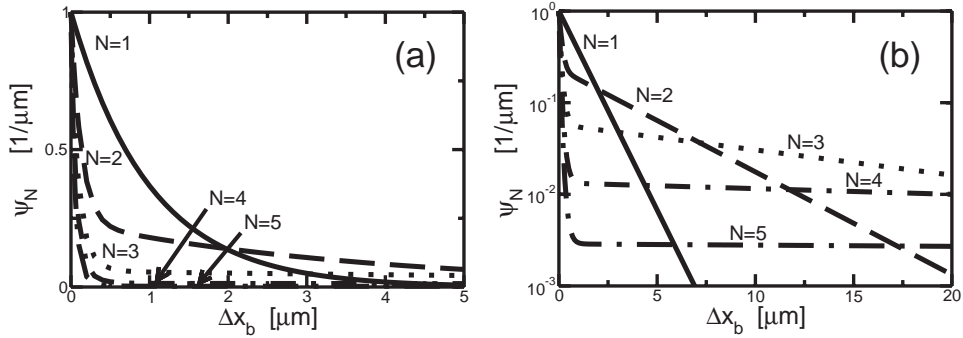}
     \caption{Probability distribution $\psi_N$ of 
       walking distance $\Delta x_\bd$ 
       for cargo particles pulled by $N=1,2,3,4,5$ kinesin motors with
       parameters as given in Table \ref{tab:param}. The same
       distributions are plotted on a linear scale in (a) and on a
       semi-logarithmic scale in (b).}
     \label{fig:verteilungen}
   \end{center}
\end{figure}

Our results for the walking distance distributions of
kinesin-pulled cargoes are shown in \fig{fig:verteilungen}. With
increasing motor number $N$, the slope of the distribution becomes
increasingly steep for small walking distances, but the distribution
becomes flatter and flatter for large walking distances.  For more
than 3 kinesins, the distribution is nearly constant for walking
distances between 5 and 20 $\mu$m, see \fig{fig:verteilungen}.

If the motors are densely packed onto the cargo particle, exclusion
effects \cite{Lipowsky__Nieuwenhuizen2001,Klumpp__Lipowsky2005} modify
the rates (\ref{eq:rate_NoWW}) as shown  in part B and Fig. 6 of the
Supporting Information. 
 For typical motor numbers $N\lesssim 10$, the effect
of exclusion on the velocity and the average walking distance is
rather small. For very dense packing, however, a reduction of the
velocity to about 35\% of the value without exclusion is obtained, in
agreement with experimental results \cite{Boehm__Unger2000a}.

{\bf Movement against  external load force. }  Let us now
consider cargo transport against a constant external force that could
be applied, e.g., by optical tweezers or other single-molecule
manipulation techniques.  This force is shared equally between the $n$ bound
motors and induces an effective interaction of the motors, since, via
the force-dependence, the transport parameters of the motors now
depend on the presence of the other motors.

The velocity of a single motor decreases essentially linearly with the
force imposed against the motor movement
\cite{Svoboda_Block1994,Hunt__Howard1994,Kojima__Yanagida1997,Visscher__Block1999}.
We therefore use the linear force--velocity relation
\begin{equation}
  \label{eq:f-v}
  v_n(F)=v\left(1-\frac{F}{n F_s}\right)
\end{equation}
for $0\leq F\leq n F_s$ and take the velocity to be constant with
$v_n(F)=v$ for $F<0$ and $v_n(F)=0$  for $F>F_s$, compare
\cite{Block__Lang2003}. The force scale $F_s$ is given by the stall
force at which a single motor stalls and stops moving.  For kinesin,
stall forces of $F_s\simeq$ 5--7 pN have been reported
\cite{Svoboda_Block1994,Hunt__Howard1994,Kojima__Yanagida1997,Visscher__Block1999}.
In the following, we use the typical value $F_s\simeq 6$pN.

The force dependence of the unbinding rates $\epsilon_n$  is given by
\begin{equation}
  \label{eq:f-eps}
  \epsilon_n(F)= n \,  \epsilon \, \exp\left(\frac{F}{n F_d}\right).
\end{equation} 
as obtained from the measurements of the walking distance of a 
single motor as a function of load \cite{Schnitzer__Block2000} in 
agreement with Kramers rate theory \cite{Kramers1940}. 
The detachment force $F_d$, which sets the force
scale here, is, in general, not equal to the stall force, although
both can be expected to have the same order of magnitude. The force
scale $F_d$ may be expressed as $F_d \equiv k_{\rm B}T/d$ which  depends on the thermal energy $k_{\rm B}T$ and on the 
extension $d$ of the potential barrier between the bound and unbound state.
For kinesins, the length scale $d$ has been reported to be $d\simeq 1.3$
nm, so that the detachment force is $F_d\simeq 3$ pN
\cite{Schnitzer__Block2000}.  

It is more difficult to estimate the  force dependence of the  binding 
rates $\pi_n$ since there are no 
experimental data about this dependence. An external load force should 
lead to a decrease of the  binding rate $\pi_0$ from the unbound state  
but this binding rate does not affect 
the  properties of the bound motor.  The binding rates $\pi_n$ 
with $n \ge 1$, on the other hand, are   expected to depend only weakly
on $F$.  This is because a pulling motor, that is subject to a certain 
strain arising from $F$, will   relax this  strain as soon as it becomes unbound
and  will then rebind from such a relaxed state.  
In other words:  unbinding and rebinding occur along 
different reaction coordinates, i.e., along
different paths in configuration space. 
Therefore,  we  take the binding rates $\pi_n$ with $n \ge 1$ to be
force-independent,
so that $\pi_n=(N-n)\piad$ for $n \ge 1$ as before.  In \eqs{eq:f-v} and
(\ref{eq:f-eps}), $v$ and $\epsilon$ are the velocity and unbinding
rate of a single motor in the absence of load, in agreement with
\eq{eq:rate_NoWW}.  A similar type of binding/unbinding dynamics but
without the active movement in the bound state  arises for the
forced rupture of adhesion molecule clusters
\cite{Bell1978,Seifert2000,Erdmann_Schwarz2004}.\footnote{In the latter 
situation, the initial state is typically given by $n=N$
  rather than by $n=1$.}

\begin{figure}[hp]
   \begin{center}
     \vspace*{0.5cm}
     \leavevmode
     \includegraphics[angle=0,width=0.9\columnwidth]{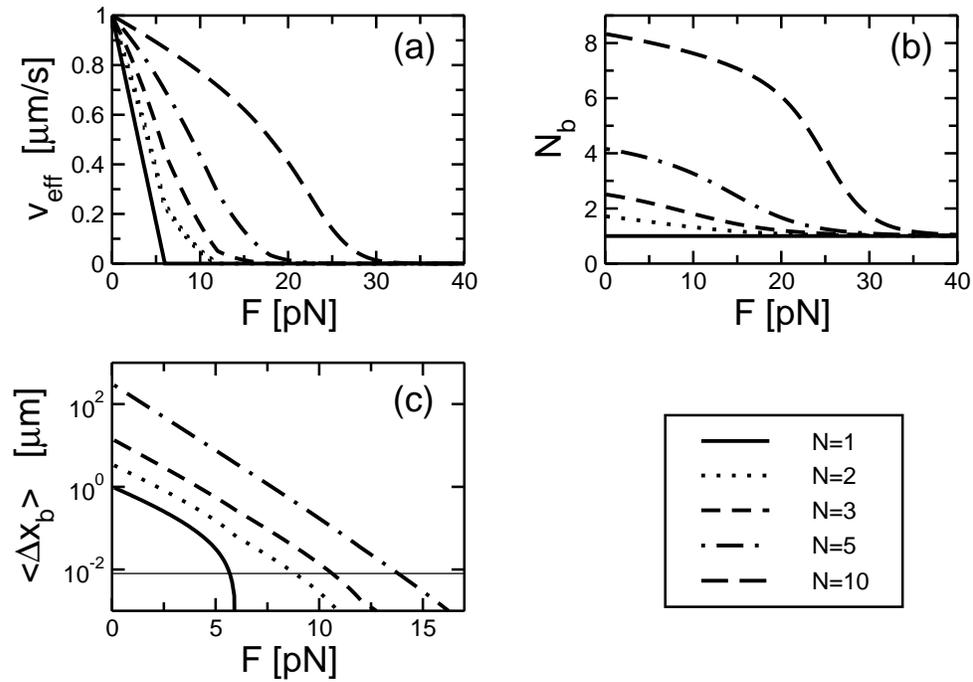}
     \caption{Transport properties of cargo particles pulled  by up to $N$ motors
       against a constant external load force $F$: (a) Average velocity
       $v_{\rm eff}$; (b) Average number $N_\bd$ of bound motors; and (c)
       Average walking distance $\langle{\Delta x_\bd}\rangle$. The chosen
       parameter values are for kinesin as in Table \ref{tab:param}. The
       horizontal line in (c) indicates the step size of 8nm. For forces for
       which $\langle{\Delta x_\bd}\rangle$ becomes comparable to or 
       smaller than the step
       size, the motors  become unprocessive.}  
     \label{fig:v_f}
   \end{center}
\end{figure}

The force--velocity relationships for cargo particles pulled by $N$
motors are shown in \fig{fig:v_f}(a).
Even though the force--velocity curve is linear for a single motor, it is
\emph{non-linear} for $N>1$, an effect that arises from the
force-dependence of the unbinding rate which implies that the average
number of bound motors decreases with increasing force, see
\fig{fig:v_f}(b).  At high forces, a cargo particle is most likely
bound to the filament by a single motor and this single motor then has
a high unbinding rate, because it is pulled off from the filament by
the total force.\footnote{For high forces with $F\protect\simg 25$pN,
  the effective unbinding rate, is given by $\epsilon_{\rm eff}\approx
  \epsilon_1(F)=\epsilon e^{F/F_d}$, independent of the number of
  motors.}  For $N>2$, the velocity decreases quickly for small and
intermediate forces, but approaches zero rather slowly for forces
close to the stall force. Indeed,  the actual stall
force for a cargo particle pulled by $N$ motors is equal to $N$ 
times the stall force $F_s$ of a single motor, but the cargo movement will 
become undetectable already at much smaller forces.

\begin{figure}[hp]
   \begin{center}
     \vspace*{0.5cm}
     \leavevmode
     \includegraphics[angle=0,width=0.9\columnwidth]{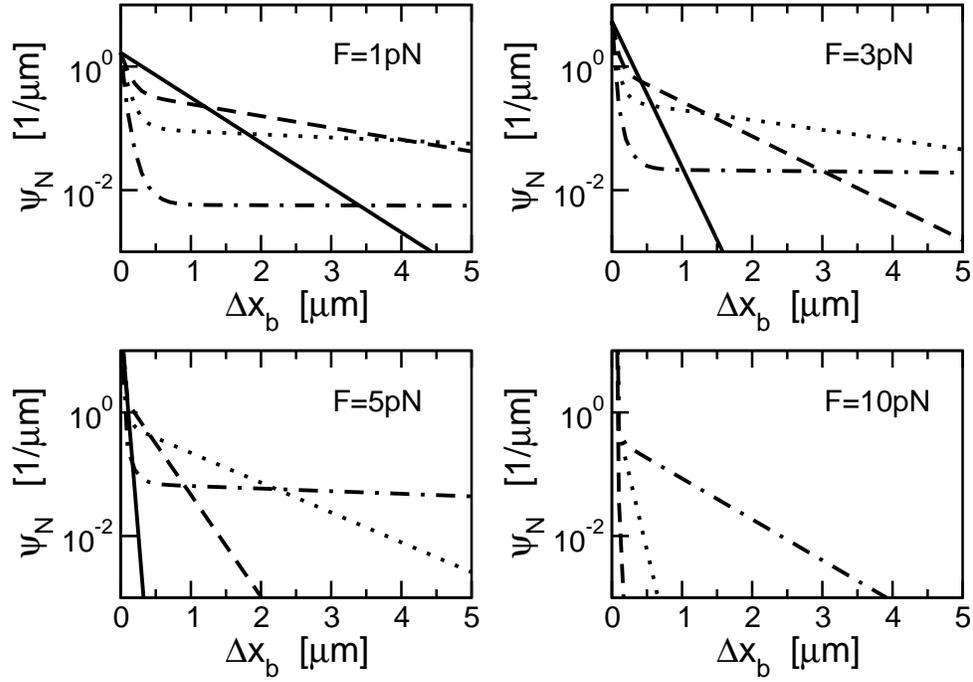}
     \caption{Probability distribution $\psi_N$ of  walking distance 
       $\Delta x_\bd$  for cargo particles pulled by up to $N=1$ (solid lines), 
       2 (dashed lines), 3 (dotted lines), and 5 (dash-dotted lines) 
       motors  against an external load force $F$.
       For $F=10$~pN, cargoes pulled by a single motor do not move, and the
       distribution $\psi_N$ contains a delta function at $\Delta x_\bd=0$.
       The distributions are plotted on a semi-logarithmic scale as in 
       \fig{fig:verteilungen}(b).}
     \label{fig:distrDXB_F}
   \end{center}
\end{figure}

The force-dependent increase of the unbinding rate is also reflected
in the corresponding decrease of the average walking distance which
is approximately exponential with increasing force $F$ for $N\geq 2$
as shown in \fig{fig:v_f}(c).\footnote{For $N=1$, the walking distance
  is given by $\langle\Delta
  x_\bd(F)\rangle=(v/\epsilon) (1-F/F_s)\,e^{-F/F_d}$.}  For very
strong forces which exceed a critical force $F_c$, the average walking
distance  becomes  comparable to the motor step size $\ell$ and the
motors become unprocessive. This critical force can be estimated 
from the implicit equation 
$\langle\Delta x_\bd(F_c)\rangle=\ell$. For kinesin which has a  step size
$\ell=8$nm, we obtain $F_c=$5.7, 8.8, 10.6, and 13.8 pN for particles
pulled  by $N=$1,2,3, and 5 motors, respectively. For $N\geq 2$, these
values are considerably smaller than the corresponding stall forces.
Force-dependent distributions of the walking distances are
shown in \fig{fig:distrDXB_F}.

\begin{figure}[hp]
   \begin{center}
     \leavevmode
     \includegraphics[angle=0,width=.9\columnwidth]{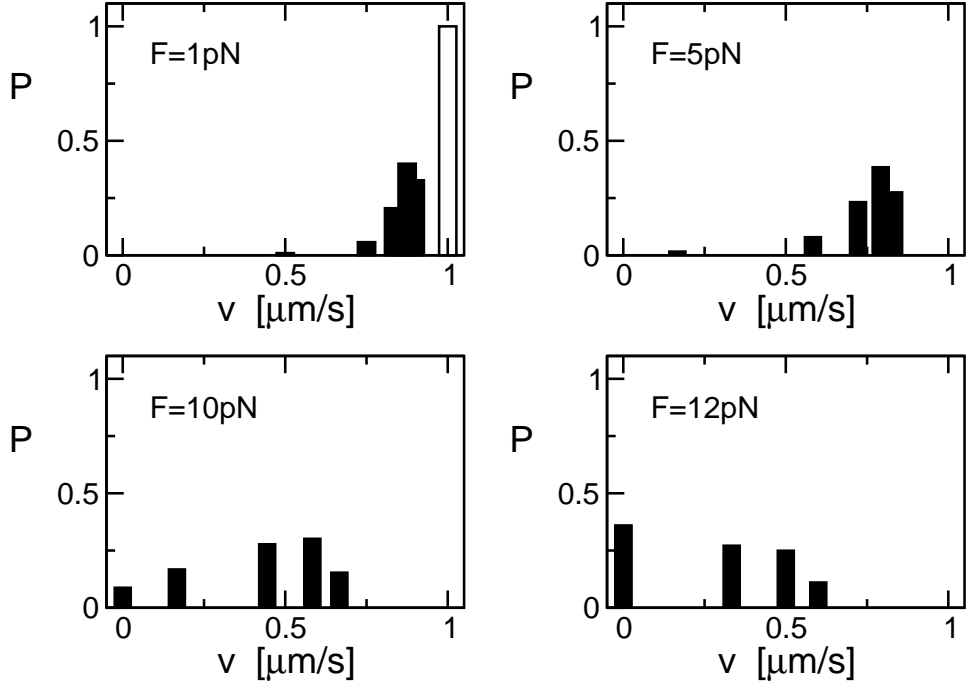}
     \caption{Probability distritubion $P$ of the instantaneous velocity 
       $v$ for 
      cargo particles that are  pulled
      by $N=5$ motors  in the presence of an external load 
      force $F$. The chosen parameter values are given in 
      Table \ref{tab:param} and apply to kinesin. 
      The white bar in the graph for $F=1$ pN indicates the distribution 
      for vanishing external force. For $F=10$ pN and $F=12$ pN, the 
      particles are stalled with a non-zero probability $P(v=0)$, because 
      no movement occurs if the cargo particle
      is pulled by a single motor for $F=10$ pN or by one or two 
      motors for  $F=12$ pN. }
     \label{fig:vHisto}
   \end{center}
\end{figure}

In the presence of an external load force, the velocity depends on the
number of motors which pull the cargo. This implies that the velocity
of such a cargo particle is switched stochastically when a motor binds
to or unbinds from the filament. The trajectory of such a cargo
therefore consists of segments with constant velocity as has 
been observed recently for vesicles dragged through the cytoplasm
\cite{Hill__Holzwarth2004,Kural__Selvin2005}. The distribution of
these velocities is shown in \fig{fig:vHisto}. With increasing load force,
the observed velocities decrease, but in addition, 
 the velocity  distribution $P(v)$  becomes broader and develops
 several peaks. The latter feature is again consistent with the 
 in vivo experiments  in Refs. \cite{Hill__Holzwarth2004,Kural__Selvin2005}.

\section{Discussion and applications}

We have presented a theoretical study of the transport properties of
cargo particles which are pulled by several
molecular motors in a cooperative fashion. Let us now discuss some applications of our results
to cellular systems.

The most prominent example for long-range transport over distances
which by far exceed the walking distances of single motors is the
transport in axons \cite{Goldstein_Yang2000}.  The cargo particles
which belong to the {\it slow} transport component such as
neurofilaments exhibit alternating periods of directed movement with
velocities of  the order of $1 \mu{\rm m}/$s and pausing periods where
essentially no movement can be detected, so that their effective
velocity is of the order of $\sim 10^{-3}-10^{-2}\mu{\rm m}/$s or
$\sim 0.1-1$ mm per day. The walking distances of the active movements
are typically a few microns, see
\cite{Wang__Brown2000,Shah_Cleveland2002}. These observations are
consistent with the assumption that these slow cargoes are transported
by one or two motors.  On the other hand, cargo particles of {\it
  fast} axonal transport such as vesicles move with velocities of $\sim
1\mu{\rm m}/$s over distances of at least centimeters. Using
\eq{eq:deltax_N_noWW}, we can estimate that the cooperation of 7--8
kinesin motors is sufficient for a walking distance in the centimeter
range. A walking distance of $\sim 1$m as necessary in the longest
axons is obtained if 10 motors drive the movement.

Our theory also gives a quantitative explanation for the effect of
microtubule-associated proteins (MAPs) such as the tau protein on the
processivity of cargo particles. 
On the one hand, the presence of tau reduces the
binding rate of kinesin to microtubules in single-molecule
experiments, but has no effect on
the velocity and walking distance of the bound kinesins 
\cite{Seitz__Mandelkow2002}. On the other hand, the movements of 
vesicles in cells transfected with tau exhibit reduced 
walking distances \cite{Trinczek__Mandelkow1999}. 
It has been proposed \cite{Seitz__Mandelkow2002} that these apparently
contradictory experimental findings can be reconciled if the vesicles
with reduced walking distance were transported by several motors. Our 
theory supports this idea, since 
\eq{eq:deltax_N} implies that the walking distance of
a cargo particle pulled  by more than one motor is affected by changes in the
binding rate. At a ratio of two tau molecules per tubulin dimer, the
binding rate of a single kinesin molecule is reduced to about 50 percent of
its value in the absence of tau \cite{Seitz__Mandelkow2002}. For
cargoes pulled  by 2, 3, and 4 kinesin motors, this reduction of the
binding rate implies a reduction of the walking distance to 64,
40, and 16 percent of the corresponding value in the absence of tau,
respectively. 

Finally, we have calculated the transport properties of cargo
particles pulled  by several motors  against an external
load force. This situation is accessible to in vitro experiments,
using, for example, bead assays and optical traps which exert constant
forces. For such experiments, our theory makes quantitative predictions
about the force-velocity relationships, the walking distances, and the
distribution of the instantaneous cargo velocities.

In addition, our theory can be applied to the movement of large
organelles in cells which experience  viscous forces of a few piconewtons
comparable to the stall force of a single motor
\cite{Luby-Phelps2000}. 
If the cargo particle moves with velocity $v_n$, it experiences the 
Stokes force $F_n=\gamma v_n$ where $\gamma$ is the corresponding 
friction coefficient. In the presence of such a force, our relation 
(\ref{eq:f-v}) leads to
\begin{equation}\label{eq:v_viscForce}
  v_n=\frac{v}{1+\gamma v/(nF_s)}\approx n \frac{F_s}{\gamma},
\end{equation}
where the asymptotic equality applies to large friction coefficients
$\gamma$.  For such a situation, two groups
\cite{Hill__Holzwarth2004,Kural__Selvin2005} have recently measured the
distribution of the instantaneous velocities as given by
\eq{eq:histv}. They found that the vesicles switch between different
values of the velocity which are peaked at integer multiples of the
smallest observed velocity.  If the friction coefficient is large
compared to $n F_s/v$ such a linear behavior is indeed predicted by
\eq{eq:v_viscForce}.

In summary, we have presented a theoretical study of the cooperative
transport of cargo particles that are pulled by up to $N$ molecular motors. We have
determined the transport properties of these cargo particles such as
their effective velocity and average walking distance (or run length). 
The latter quantity is strongly affected by the maximal number $N$ of pulling
motors, and the cooperation of several motors enables efficient
transport over large distances. Our approach provides a quantitative 
theoretical
basis for the interpretation of a number of recent experiments and
makes quantitative predictions which can be tested experimentally. 
The theoretical framework introduced here
can  be extended to more complex situations such as the 
transport of   cargo particles that are attached to several species of motors.  These 
 different species may have different velocities or may even move in 
 opposite directions. Likewise, our theory
can be extended to load forces that depend on the displacement of the cargo 
particle or change with 
time. A relatively simple example for  such a variable load force is provided by 
laser traps,  which are used in motility assays in order to exert  harmonic 
force potentials for small particle displacements. More complex examples are
found for  the cytoskeletal transport in biological cells, where the cargo particle is 
pulled  through a meshwork of membranes and filaments that can act as 
steric barriers or adhesive surfaces and, thus, can exert various types of
position-dependent forces on this particle.

\begin{acknowledgments}
  The authors thank Janina Beeg, Melanie M\"uller, Thorsten Erdmann,
  and Ulrich Schwarz for stimulating discussions.
\end{acknowledgments}


\newpage

\begin{table}[h]
\begin{center}
\begin{tabular}{l||c|c|c}
Parameter & Symbol & Value for kinesin & Reference \\
\hline\hline
Velocity & $v$ & 1 $\mu$m/s & \cite{Block__Schnapp1990,Vale__Yanagida1996} \\
Unbinding rate & $\epsilon$ & 1/s & \cite{Block__Schnapp1990,Vale__Yanagida1996}\\
Binding rate & $\piad$ & 5/s & \cite{Leduc__Prost2004}\\
Stall force & $F_s$ & 6pN & \cite{Visscher__Block1999,Schnitzer__Block2000} \\
Detachment force & $F_d$ & 3pN & \cite{Schnitzer__Block2000}\\
\end{tabular}
\end{center}
\caption{Model parameters for single motors and values for conventional 
kinesin.}\label{tab:param}
\end{table}

\vspace*{5cm}

\newpage

\renewcommand{\theequation}{S.\arabic{equation}}

\newpage
{\huge Supporting text}

\appendix
\section*{A.$\quad$Technical Aspects of the Calculations}

\subsection*{A.1.$\quad$Mean First Passage Times}

The effective unbinding rate as given by Eq.\  {\bf 8} 
has been derived by a simple equilibrium argument.  The same equation 
can also be obtained by calculating the mean
first passage time. Let us denote by $T_{m,N}$ the mean first 
passage time to the
state $|0\rangle$ with no bound motor if we start from state 
$|m\rangle$ with $m$ bound 
motors at time
$t=0$ (the second index $N$ indicates the total number of
motors). 
The effective unbinding rate is then given by $1/T_{1,N}$,
since the cargo particle first binds to the filament through a single
motor.

The first passage times fulfill the  recursion relations
\begin{equation}
T_{m,N}=\frac{1}{\epsilon_m+\pi_m}+\frac{\pi_m}{\epsilon_m+\pi_m}
T_{m+1,N}+\frac{\epsilon_m}{\epsilon_m+\pi_m}T_{m-1,N}
\end{equation}
for $m\neq 0, N$, , see, e.g., ref.\ 1, 
with the boundary recursions
\begin{eqnarray}
  T_{N,N} & = & \frac{1}{\epsilon_N}+T_{N-1,N} \quad {\rm and}\\
  T_{0,N} & = & 0.
\end{eqnarray}
Because of the boundary condition   $T_{0,N}  =  0$, the recursion 
relation Eq.\  {\bf S.1} with $m=1$ can be used to express $T_{2,N}$ in terms of
$T_{1,N}$. Next, starting from Eq.\  {\bf S.1} with $m=2$ and using the relation 
between  $T_{2,N}$ and  $T_{1,N}$, we can also express
$T_{3,N}$ in terms of $T_{1,N}$.  Iteration of this procedure leads to 
explicit expressions for $T_{m,N}$ in terms of $T_{1,N}$. 
Finally, when these expressions are used in  Eq.\  {\bf S.2}, we obtain an 
implicit equation for  $T_{1,N}$, which is solved by
\begin{equation}
  T_{1,N}=\frac{1}{\epsilon_1}\left(1+\sum_{i=1}^{N-1}
  \prod_{n=1}^{i}\frac{\pi_n}{\epsilon_{n+1}}\right),
\end{equation}
which is exactly the inverse of Eq.\ {\bf 8}.

\subsection*{A.2.$\quad$Distribution of Unbinding Times}

To calculate the distribution of unbinding times, we consider the
probability distribution for the passage from state $|m\rangle$ with $m$ bound
motors at time $t=0$ to the unbound state $|0\rangle$ at time $t$, which we denote
by $\tilde\psi_{m,N}(t)$. The distribution of unbinding times is then
given by $\tilde\psi_N(\Delta t_\bd)\equiv\tilde\psi_{1,N}(t=\Delta
t_\bd)$ since the initial bound state of the cargo particle is
provided by  state $|1\rangle$ with $m=1$ for which the particle is bound to the
filament by a single motor.

The probability distributions $\tilde\psi_{m,N}(t)$ fulfill the
recursion relations
\begin{equation}
  \tilde\psi_{m,N}(t)=\int_0^t e^{-(\epsilon_m+\pi_m)\tau}\left[\pi_m\,\tilde\psi_{m+1,N}
  (t-\tau)+\epsilon_m\,\tilde\psi_{m-1,N}(t-\tau)\right] \dd\tau
\end{equation}
for $m\neq 0,N$,
\begin{eqnarray}
  \tilde\psi_{N,N}(t) & = & \int_0^t e^{-\epsilon_N\tau}\epsilon_N\,\tilde\psi_{N-1,N}(t-\tau)
   \dd\tau,\qquad{\rm and}\\
  \tilde\psi_{0,N}(t) & = & \delta(t).
\end{eqnarray}
These recursion relations are obtained by considering the first
binding/un\-binding event explicitly, summing over the two
possibilities for this step (to $m\pm 1$), and integrating over all
possible times $\tau$ at which this first event occurs.  The
exponential terms express the probability that no binding/unbinding
event occurred until the time $\tau$.

Using Laplace transforms, we can transform the convolution integrals
into algebraic equations and iteratively obtain all the Laplace
transformed distributions $\tilde\psi_{m,N}(s)$.  The solution is
given by a finite continued fraction of depth $N$, which has the form
\begin{equation}\label{eq:cont_frac}
  \tilde\psi_{1,N}(s)=\frac{\epsilon_1}{\epsilon_1+s+\pi_1\left(1-\frac{\epsilon_2}
  {\epsilon_2+s+\pi_2\left(1-\frac{...}{...+\pi_{N-1}\left(1-   \frac{\epsilon_N}
{\epsilon_N+s}\right)}...\right)}\right)},
\end{equation}
see chapter 9 of ref.\ 2. 

In general, the inverse Laplace transform of Eq.\ {\bf S.8} 
can be expressed as
\begin{equation}
  \tilde\psi_{1,N}(t)=\sum_{i=1}^{N}e^{-z_i t}\ {\rm Res}(-z_i),
\end{equation}
where the parameters $-z_i$ are the poles of $\tilde\psi_{1,N}(s)$ and
${\rm Res}(-z_i)$ are the corresponding residues (see ref.\ 3).  
All poles $-z_i$ are real and negative. Using the
definition $\tilde\psi_N(\Delta t_\bd)\equiv \tilde\psi_{1,N}(t=\Delta
t_\bd)$ in the relation Eq.\ {\bf S.9}, we obtain the binding time distribution
as given by Eq.\ {\bf 9}. 

In general, the poles and the residues have to be calculated
numerically, but in the two simplest cases, $N=1$ and $N=2$, the
inverse Laplace transform can be obtained in closed form. For $N=1$,
we can check that we recover the single exponential
$\tilde\psi_{1,1}(t)=\epsilon_1 e^{-\epsilon_1 t}$, and for $N=2$ the
first passage time distribution is given by
\begin{eqnarray}
  \tilde\psi_{1,2}(t) & = & \frac{\epsilon}{2}\Big[ \Big(1-\frac{\epsilon_1+\pi_1-\epsilon_2}
{R}\Big) e^{-\frac{1}{2}(\epsilon_1+\epsilon_2+\pi_1-R)t}\nonumber\\
 &  & {}+\Big(1+\frac{\epsilon_1+\pi_1-\epsilon_2}{R}\Big) e^{-\frac{1}{2}
(\epsilon_1+\epsilon_2+\pi_1+R)t}\Big]
\end{eqnarray}
with
$R\equiv\sqrt{(\epsilon_1+\epsilon_2+\pi_1)^2-4\epsilon_1\epsilon_2}$.

\section*{B$\quad$Mutual Exclusion of Motors}

In general, several motor molecules, which are bound to a certain
cargo particle, may compete for the same binding site of the filament.
Such a competition may arise, for example, because the motor molecules
are densely packed on the cargo particle or because they move along a
single protofilament of the microtubule. In such a situation, mutual
exclusion or hard core repulsion between the motors should to be taken
into account. Exclusion reduces the binding of motors to the filament
and the velocity of the bound motors (4,5).  
Within a
mean-field approximation, these two effects can be incorporated into
our model by using modified binding rates $\pi_n$ and modified bound state velocities
$v_n$ as given by
\begin{equation}
  \label{eq:rate_HC}
  \pi_n = (N-n)\piad\left[1-\frac{n}{N_s}\right] \qquad{\rm and}\qquad
  v_n = v\left[1-\frac{n-1}{N_s-1}\right]
\end{equation}
for $ n \leq N_s$ where $N_s$ is the number of accessible binding sites
that the motors can reach for a given position of the cargo particle.
The terms $[1-n/N_s]$ and $[1-(n-1)/(N_s-1)]$ describe the probability
that the site to which a motor attempts to bind or to move is not
occupied by another motor.\footnote{The difference between these two
  expressions arises from a finite-size effect. When an unbound motor
  attempts to bind to the state $|n\rangle$, it encounters 
   $n$ out of $N_s$ binding sites that are already occupied. 
  In contrast,  a bound motor in state $|n\rangle$ `feels' only  
  $n-1$ motors, which are bound to
  $n-1$ out of  the remaining $N_s-1$ binding sites.} 
  For $n \geq N_s$, all binding
  sites that could be reached by the motors are occupied, so that
$\pi_n=0$ for $n \geq N_s$.  The unbinding rates $\epsilon_n$ are
unaffected by exclusion and are again given by $\epsilon_n = n \, \epsilon$.
If the number of accessible binding sites $N_s$ is much larger than
the number of motors attached to the cargo particle, the motors are
effectively noninteracting, and \eq{eq:rate_HC} can be approximated
by Eq.\ {\bf12}.

For typical cargoes such as beads or vesicles with diameters between
100~nm and 1~$\mu$m, we can estimate the number of binding sites
within the contact zone of the cargo particle to be of the order of
50--150, while the number of motors is typically 1--10.  For these
motor numbers and for the parameter values corresponding to kinesin,
exclusion effects are rather small. Inspection of
Fig.\ 6{\it a}  shows that the average velocity is reduced by
a few percent as compared to noninteracting motors.  The average
walking distance is more sensitive to exclusion, but still of the same
order of magnitude as for noninteracting motors. For example, for
$N=5$, the walking distance which is $\simeq$ 310~nm without exclusion
is reduced to 280~nm, see Fig.\ 6{\it b}.

If motors are closely packed on the cargo particle, i.e.\ for $N\simeq
N_s$, a reduction of the velocity to about 35 percent of the value
without exclusion is obtained as shown in Fig.\ 6{\it a}.  For
very high motor densities, a reduction of the velocity of the order of
50 percent has indeed been observed both in microtubule gliding assays (6) 
and bead assays (J. Beeg, private
communication) for kinesin.

In principle, exclusion implies that the walking distance exhibits a
maximum as a function of the number of motors, since at very large
motor numbers, the velocity approaches zero. Using the rates given by 
Eq.\ {\bf S.11}   
in Eq.\ {\bf 11}, we find, however, that this maximum
occurs at walking distances that are far too large to be
experimentally accessible.\\[.5cm]


\noindent
1.\ van Kampen, N. G. (1992)  \newblock {\em Stochastic
    Processes in Physics and Chemistry} (Elsevier, Amsterdam).\\

\noindent      
2.\ Risken, H. (1989) \newblock {\em The Fokker--Planck
    Equation} (Springer, Berlin) 2nd Ed.\\

\noindent  
3.\ Schiff, J.~L. (1999) \newblock {\em The Laplace
    Transform: Theory and Applications} (Springer, New York). \\
      
\noindent
4.\ Lipowsky, R., Klumpp, S. \& Nieuwenhuizen, 
T.~M. (2001)  \newblock {\em Phys.\ Rev.\ 
    Lett.} {\bf 87}, 108101.\\
    
\noindent
5.\ Klumpp, S., Nieuwenhuizen, T.~M. \&
  Lipowsky, R. (2005) \newblock {\em
    Biophys.\ J.} {\bf 88}, 3118--3132.\\
 
\noindent 
6.\ B\"ohm, K.~J., Stracke, R. \& Unger, E. (2000)
  \newblock {\em Cell Biol.\ Int.} {\bf 24}, 335--341.


\begin{figure}[b]
   \begin{center}
     \leavevmode
     \includegraphics[angle=0,width=.9\columnwidth]{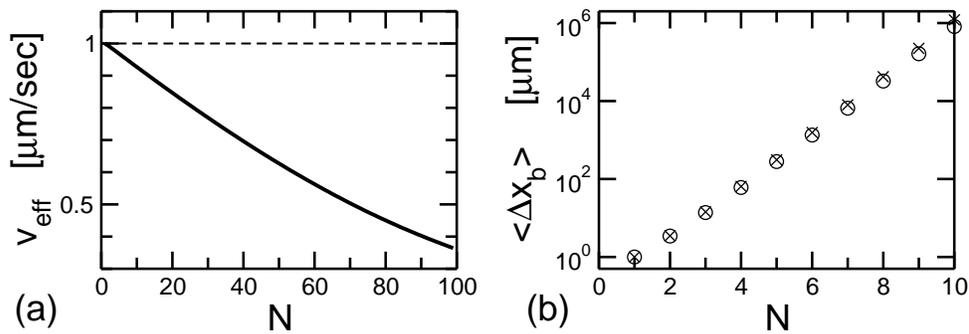}
     \caption{Exclusion effects: (a) Average velocity $v_{\rm eff}$ and (b) 
       average walking distance $\langle\Delta x_\bd\rangle$ as
       functions of the number $N$ of motors attached to the cargo.
       The chosen parameter values are those of kinesin as described
       in the text.  The number of binding sites which are accessible
       to the motors for a given position of the cargo particle is
       $N_s=100$ as appropriate for a cargo with radius $\sim 1\mu$m
       [solid line in (a) and circles in (b)]. The values indicated by
       the dashed line in (a) and the crosses in (b) are obtained if
       exclusion effects are not taken into account.  For typical
       motor numbers $N \protect\siml 10$, direct comparison shows
       that exclusion effects are small.}
     \label{fig:exclusion}
   \end{center}
\end{figure}

\end{document}